\documentclass[journal]{IEEEtran}
\ifCLASSINFOpdf
\else
\fi
\usepackage{color, colortbl}
\usepackage{graphicx}
\usepackage{amsmath}
\usepackage{tabularx}
\usepackage{multirow}
\usepackage{booktabs}
\usepackage{cite}
\usepackage{datetime}
\usepackage{amssymb}
\usepackage{float}
\usepackage{stfloats}
\usepackage{lipsum}
\usepackage{mathtools}
\usepackage{arydshln}
\usepackage{stfloats}
\usepackage{amsthm}

\usepackage{cuted}

\newcommand{\md}{\mathcal{D}}

\newcommand{\diff}[2]{\frac{\mathrm{d}{#1}}{\mathrm{d}{#2}}}
\newcommand{\pdiff}[2]{\frac{\partial{#1}}{\partial{#2}}}
\newcommand{\x}{\hspace{2.5mm}}

\definecolor{Pink}{rgb}{0.86, 0.19, 0.48}
\definecolor{LightBlue}{rgb}{0.9,0.9,1.0}
\definecolor{LightRed}{rgb}{1.0,0.9,0.9}

\newcommand{\parcommand}[1]{\textcolor{Pink}{\textit{#1}}}

\theoremstyle{remark}
\newtheorem{remark}{Remark}

\begin{document}
\title{Contingency Analysis Based on Partitioned and Parallel Holomorphic Embedding}

\author{Rui~Yao,~\IEEEmembership{Senior~Member,~IEEE,}~Feng~Qiu,~\IEEEmembership{Senior~Member,~IEEE,}~and~Kai~Sun,~\IEEEmembership{Senior~Member,~IEEE}% <-this % stops a space
\thanks{This work was supported by the Advanced Grid Modeling program of U.S. Department of Energy.}
\thanks{R. Yao and F. Qiu are with the Division of Energy Systems, Argonne National Laboratory, Lemont 60439, USA. (emails: ryao@anl.gov, fqiu@anl.gov).}
\thanks{K. Sun is with the Department of EECS, the University of Tennessee, Knoxville, TN 37996, USA (email: kaisun@utk.edu).}}


\markboth{%\textbullet~\textbullet~\textbullet~\textbullet~\textbullet~Z~\textbullet
}%
{Shell \MakeLowercase{\textit{et al.}}: }



\maketitle

\begin{abstract}
In the steady-state contingency analysis, the traditional Newton-Raphson method suffers from non-convergence issues when solving post-outage power flow problems, which hinders the integrity and accuracy of security assessment. In this paper, we propose a novel robust contingency analysis approach based on holomorphic embedding (HE). The HE-based simulator guarantees convergence if the true power flow solution exists, which is desirable because it avoids the influence of numerical issues and provides a credible security assessment conclusion. In addition, based on the multi-area characteristics of real-world power systems, a partitioned HE (PHE) method is proposed with an interface-based partitioning of HE formulation. The PHE method does not undermine the numerical robustness of HE and significantly reduces the computation burden in large-scale contingency analysis. The PHE method is further enhanced by parallel or distributed computation to become parallel PHE (P${}^\mathrm{2}$HE). Tests on a 458-bus system, a synthetic 419-bus system and a large-scale 21447-bus system demonstrate the advantages of the proposed methods in robustness and efficiency.
\end{abstract}


% \fi
\IEEEpeerreviewmaketitle

\section{Introduction}
\IEEEPARstart{C}{ontingencies} may pose risks to power system operations and thus it is necessary to perform contingency analysis. Steady-state contingency analysis \cite{ejebe1995methods} screens and simulates large numbers of single or multiple outages based on power flow analysis, which is a routine computation task in power system planning and operations. The Newton-Raphson (NR) method is commonly used to solve AC power flow problems. However, the NR method is sensitive to the initial solution, it frequently fails to converge in contingency analysis. And it has long been a complaint from the industry that people cannot infer whether the convergence failure is caused by the loss of a power flow solution (system collapse) or by numerical issues \cite{nerc2018irol,dong2012dealing}, which significantly affects the integrity and credibility of security assessment. Continuation methods \cite{flueck2000new, yao2012static} were proposed to improve the numerical robustness of NR method, and they can also be used for contingency analysis by applying continuation on the outage component parameters. However, because continuation methods need to invoke the NR method repeatedly, their efficiency cannot satisfy industrial demands. In recent years, holomorphic embedding (HE) methods have emerged as a promising approach to reliably solve the nonlinear steady-state and dynamic problems in power systems \cite{trias2012holomorphic,yao2018voltage,liu2017online}. Unlike the Newton-Raphson method which depends on search and trial, the HE approach approximates the system solutions using high-order power series or other analytical forms\cite{rao2015holomorphic}. HE theoretically guarantees convergence if the solution exists \cite{stahl1997convergence} and also has excellent numerical performance \cite{rao2018theoretical} which is promising for power system analysis. The guaranteed convergence is also very desirable for the contingency analysis. This paper will propose HE approaches for efficient and robust contingency analysis. The basic HE formulation and algorithm for solving contingency analysis based on HE will be first presented.

The major computational performance bottleneck of HE in contingency analysis is on solving linear equations at the scale of system size. When performing contingency analysis on very large-scale systems, the computation still could be intensive. A power system usually has hierarchical structures \cite{cao2017coordinating}: The buses and branches on the highest voltage levels constitute the bulk power system, and the sub-transmission systems and distribution systems with lower voltage levels are connected to the bulk system. With the increasing interdependency across the power system, contingencies and outages may propagate across the boundaries between the bulk power system and the lower-level systems \cite{li2015transmission}. Therefore, it is necessary to perform contingency analysis for the entire system. The bulk power system usually has meshed topology due to the reliability needs, while the lower-level systems are less meshed and have very few connection points to the bulk system. Therefore, other than the cumbersome method of directly solving the contingency problems on the whole system, it would be more efficient to decompose the system and perform the analysis on each part concurrently with coordination on the boundaries. Some traditional methods separate the computation on the bulk power system and the lower-level systems, and iterate on the boundaries \cite{huang2016integrated}. However, those methods do not guarantee convergence of the iterations. Also, the iterations may be very slow to reach convergence.
To overcome the convergence and efficiency issues of contingency analysis in multi-area systems, we propose a partitioned HE (PHE) method that separates the HE models by the areas of systems and couples them with a generic voltage-current interface. Such a PHE method preserves the numerical robustness of original HE and reduces the computational burden by splitting computation on smaller lower-level systems. Furthermore, the computation on lower-level systems is independent and can be parallelized, which forms the parallel PHE (P${}^\mathrm{2}$HE) method. The PHE and P${}^\mathrm{2}$HE methods demonstrate promising numerical robustness and satisfactory efficiency on large-scale multi-area systems.

The contributions of this paper are threefold:

1) The general HE algorithm for contingency analysis is presented and some useful properties of HE that are favorable for computation and system decomposition are discovered.

2) A partitioned HE (PHE) for contingency analysis is developed, which equivalently decomposes the original HE problem into smaller problems. Complexity analysis proves that PHE reduces the computation burden of large-scale contingency analysis. In contrast to other approaches that iterate on the boundaries and have risk of divergence, the proposed PHE method does not compromise the numerical robustness. 

3) A parallel partitioned HE (P${}^\mathrm{2}$HE) is further developed based on the PHE formulation. The P${}^\mathrm{2}$HE is suitable for parallel or distributed computation to further achieve significant acceleration of HE-based contingency analysis.

The rest of the paper is organized as follows. Section II proposes the method for HE-based contingency analysis and derives important properties of the HE approach, which lays foundation for the partition. Section III proposes the PHE method and proves its advantageous efficiency. Section IV proposes P${}^\mathrm{2}$HE methods. Section V presents the test cases on a 458-bus system, a synthetic 419-bus system and a large-scale 21447-bus system. Section VI is the conclusion.

\section{HE formulation for contingency analysis}
\subsection{Algorithm of HE-based contingency analysis}
The power flow equation can be written as the following:
\begin{equation}\label{eqn:power_flow_eqn_nt_pvpq}
(P_i-jQ_i)W_i^*-\sum_{j}Y_{ij}V_j-I_{Li}=0
\end{equation}
where bus $i$ can be a PQ or PV bus. $P_i$ and $Q_i$ are the active power and reactive power injections to bus $i$, respectively. $V_i$ is the voltage of bus $i$, whose reciprocal is $W_i$, and $Y_{ij}$ is the row-$i$, column-$j$ element of the admittance matrix $\mathbf{Y}$. $\mathbf{Y}$ can also include the constant-impedance load components. $I_{Li}$ is the gross current to all other components on bus $i$. The inclusion of $I_{Li}$ has some flexibility; e.g., the injection current from a PQ load component can be counted either in the first term in (\ref{eqn:power_flow_eqn_nt_pvpq}) or in $I_{Li}$. For a PV bus, the reactive power injection $Q_i$ is unknown, but the voltage magnitude is given:
\begin{equation}\label{eqn:power_flow_eqn_nt_pvpq_v}
V_iV_i^*=|V_i^{sp}|^2,i\in S_{PV}
\end{equation}
where $S_{PV}$ is the set of PV buses.

The change of the admittance matrix causes the power flow solution to change. Assume that due to a contingency, the admittance matrix changes to $\mathbf{Y}+\Delta\mathbf{Y}$, and we can establish the holomorphic embedding (HE) formulation as follows:
\begin{equation}\label{eqn:power_flow_ctg_nt_pvpq}
\begin{aligned}
(P_i(\alpha)&-jQ_i(\alpha))W_i^*(\alpha)\\
-\sum_{l}&(Y_{ij}+\alpha\Delta Y_{ij})V_j(\alpha)-I_{Li}(\alpha)=0,i\notin S_{SL}\\
&V_i(\alpha)V_i^*(\alpha)=|V_i^{sp}|^2,i\in S_{PV}
\end{aligned}
\end{equation}
where $S_{SL}$ is the set of slack bus, and a slack bus $i$ is assumed to have a constant voltage phasor $V_i$. In (\ref{eqn:power_flow_ctg_nt_pvpq}), $\alpha=0$ corresponds to the pre-contingency state, and $\alpha=1$ corresponds to the post-contingency state. In the steady-state contingency analysis, $P_i(\alpha)$ is given as a constant $P_i$. For a PV bus, $Q_i(\alpha)$ is to be solved, while for other buses, $Q_i(\alpha)$ is given as a constant $Q_i$. HE aims to derive the solution of (\ref{eqn:power_flow_ctg_nt_pvpq}) as power series of $\alpha$, e.g.
\begin{equation}\label{eqn:example_he}
\mathbf{V}(\alpha)=\mathbf{V}[0]+\mathbf{V}[1]\alpha+\mathbf{V}[2]\alpha^2+\cdots
\end{equation}
or the corresponding Pad\'e approximations.

According to the rules for deriving HE coefficients, we can obtain the linear equations in (\ref{eqn:hem_nt_sol_linear}), where we reorder the buses so that the PV buses follow the PQ buses. $\mathbf{G}$ and $\mathbf{B}$ are real and imaginary parts of admittance matrix $\mathbf{Y}$. $\mathbf{Y}$, $\mathbf{V}$ and $\mathbf{W}$ only include PQ and PV buses, and $\mathbf{V}_{SL}$ is the voltage of the $\mathrm{V\theta}$ bus. $\mathbf{Y}_{SL}$ has rows corresponding to PQ and PV buses and a column corresponding to the $\mathrm{V\theta}$ bus. $\md(\cdot)$ stands for a diagonal matrix. $\mathbf{C}$ and $\mathbf{D}$ are the real and imaginary parts of bus voltage $\mathbf{V}$, and $\mathbf{E}$ and $\mathbf{F}$ are the real and imaginary parts of $\mathbf{W}$. $(\cdot)_{PQ}$, $(\cdot)_{PV}$ and $(\cdot)_{SL}$ stand for the variables corresponding to PQ, PV and $\mathrm{V\theta}$ buses, respectively.
\begin{equation}\label{eqn:hem_nt_sol_linear}
\begin{aligned}
&\resizebox{1\hsize}{!}{$\left[\begin{array}{c;{2pt/2pt}c;{2pt/2pt}c;{2pt/2pt}c;{2pt/2pt}c}
	-\mathbf{G} &  \mathbf{B} & \md(\mathbf{P}_0) & -\md(\mathbf{Q}_0) & \begin{matrix}\mathbf{0}\\-\md(\mathbf{F}_{PV}[0])\end{matrix} \\\hdashline[2pt/2pt]
	-\mathbf{B} & -\mathbf{G} &-\md(\mathbf{Q}_0) & -\md(\mathbf{P}_0) & \begin{matrix}\mathbf{0}\\-\md(\mathbf{E}_{PV}[0])\end{matrix} \\\hdashline[2pt/2pt]
	\md(\mathbf{E}[0]) & -\md(\mathbf{F}[0]) & \md(\mathbf{C}[0]) & -\md(\mathbf{D}[0]) & \mathbf{0} \\\hdashline[2pt/2pt]
	\md(\mathbf{F}[0]) &  \md(\mathbf{E}[0]) & \md(\mathbf{D}[0]) &  \md(\mathbf{C}[0]) & \mathbf{0}\\\hdashline[2pt/2pt]
	\begin{matrix}\mathbf{0}&\md(\mathbf{C}_{PV}[0])\end{matrix} & \begin{matrix}\mathbf{0}&\md(\mathbf{D}_{PV}[0])\end{matrix} & \mathbf{0} & \mathbf{0} & \mathbf{0}  
	\end{array}	\right]
	\left[\begin{array}{c}
	\mathbf{C}[n]\\\mathbf{D}[n]\\\mathbf{E}[n]\\\mathbf{F}[n]\\\mathbf{Q}_{PV}[n]
	\end{array}\right]$}\\
&\resizebox{1\hsize}{!}{$=\left[\begin{array}{c}
	\Re\displaystyle{\left(\mathbf{I}_{LPQ}[n]\right)}\\
	\Re\displaystyle{\left(j\sum_{k=1}^{n-1}\mathbf{Q}_{PV}[k]\circ \mathbf{W}_{PV}^*[n-k]+\mathbf{I}_{LPV}[n]\right)}\\\hdashline[2pt/2pt]
	\Im\displaystyle{\left(\mathbf{I}_{LPQ}[n]\right)}\\
	\Im\displaystyle{\left(j\sum_{k=1}^{n-1}\mathbf{Q}_{PV}[k]\circ \mathbf{W}_{PV}^*[n-k]+\mathbf{I}_{LPV}[n]\right)}\\\hdashline[2pt/2pt]
	\displaystyle{\Re\left(-\sum_{k=1}^{n-1}\mathbf{W}[k]\circ \mathbf{V}[n-k]\right)}\\\hdashline[2pt/2pt]
	\displaystyle{\Im\left(-\sum_{k=1}^{n-1}\mathbf{W}[k]\circ \mathbf{V}[n-k]\right)}\\\hdashline[2pt/2pt]
	\displaystyle{-\frac{1}{2}\sum_{k=1}^{n-1}\mathbf{V}_{PV}[k]\circ \mathbf{V}^*_{PV}[n-k]}
	\end{array}
	\right]+
	\left[\begin{array}{c}
	\Re\displaystyle{\left(\Delta\mathbf{Y}\mathbf{V}[n-1]+\Delta\mathbf{Y}_{SL}\mathbf{V}_{SL}[n-1]\right)}\\\hdashline[2pt/2pt]
	\Im\displaystyle{\left(\Delta\mathbf{Y}\mathbf{V}[n-1]+\Delta\mathbf{Y}_{SL}\mathbf{V}_{SL}[n-1]\right)}\\\hdashline[2pt/2pt]
	\mathbf{0}\\\hdashline[2pt/2pt]
	\mathbf{0}\\\hdashline[2pt/2pt]
	\mathbf{0}
	\end{array}
	\right]
	$}
\end{aligned}
\end{equation} 

The 0th-order HE coefficients are known i.e., the pre-contingency system states. Then the arbitrarily higher-order HE coefficients can be calculated recursively by solving (\ref{eqn:hem_nt_sol_linear}) \cite{yao2020novel}. A multi-stage scheme \cite{wang2017multi} is also used to expand the effective range of HE. If the computation successfully reaches $\alpha=1$, then the solution at $\alpha=1$ is the post-contingency solution. Otherwise, the system is considered as collapsed after the contingency. 
\subsection{Some Properties of Holomorphic Embedding}
\label{subsec:he_properties}
More generally, the HE approximates the solution of the following $\alpha$-parameterized system 
\begin{equation}\label{eqn:abstract_he}
\mathbf{f}(\mathbf{x}(\alpha),\alpha)=\mathbf{0}
\end{equation}
where $\mathbf{f}$ is an analytic function, $\mathbf{x}$ is $M\times 1$ vector and $\mathbf{x}(\alpha)$ has the following power-series form:
\begin{equation}\label{eqn:power_flow_solution}
\mathbf{x}(\alpha)=\mathbf{x}[0]+\mathbf{x}[1]\alpha+\mathbf{x}[2]\alpha^2+\cdots
\end{equation}

The core of HE is to derive the equations of the coefficients in (\ref{eqn:power_flow_solution}), which transforms the operations in (\ref{eqn:abstract_he}) to the relationship of the coefficients based on a set of rules, e.g.:
\begin{equation}\label{eqn:power_he_ops}
\begin{aligned}
ax(\alpha)+b&\leftrightarrow ax[n]+b\\
x(\alpha)y(\alpha)&\leftrightarrow (x*y)[n]
\end{aligned}
\end{equation}
where $*$ is convolution: $(x*y)[n]=\sum_{k=0}^{n}x[k]y[n-k]$.

For classical power system steady-state analysis, the HE formulation can usually be generalized as the following form:
\begin{equation}\label{eqn:general_he}
\resizebox{0.91\hsize}{!}{$(\mathbf{A}_0+\alpha\mathbf{A}_1)(\mathbf{x}(\alpha)\otimes\mathbf{x}(\alpha))+(\mathbf{B}_0+\alpha\mathbf{B}_1)\mathbf{x}(\alpha)+\mathbf{C}_0+\alpha\mathbf{C}_1=\mathbf{0}$}
\end{equation}
where $\otimes$ is Kronecker product, $\mathbf{x}\otimes\mathbf{x}=(\mathbf{x}\otimes\mathbf{1})\circ(\mathbf{1}\otimes\mathbf{x})$, the $\circ$ is Hardamard (element-wise) product. $\mathbf{A}_0$, $\mathbf{A}_1$, $\mathbf{B}_0$, $\mathbf{B}_1$, $\mathbf{C}_0$ and $\mathbf{C}_1$ are constant matrices/vectors. According to the rules for deriving the equations of HE coefficients, for $n$th-level terms, the equations are:
\begin{equation}\label{eqn:general_he_sol}
\begin{aligned}
&\mathbf{A}_0\left((\mathbf{x}\otimes\mathbf{1})*(\mathbf{1}\otimes\mathbf{x})\right)[n]+
\mathbf{A}_1\left((\mathbf{x}\otimes\mathbf{1})*(\mathbf{1}\otimes\mathbf{x})\right)[n-1]\\
&+\mathbf{B}_0\mathbf{x}[n]+\mathbf{B}_1\mathbf{x}[n-1]+\delta_{n,0}\mathbf{C}_0+\delta_{n,1}\mathbf{C}_1=\mathbf{0}
\end{aligned}
\end{equation}
where $\delta_{n,m}=1$ if $m=n$; otherwise $\delta_{n,m}=0$. 

Alternatively, note the following two formulas regarding high-order derivatives:
\begin{equation}\label{eqn:aux_derivative}
\begin{aligned}
&\left.\diff{^n (\alpha f(\alpha))}{\alpha^n}\right|_{\alpha=0}=nf^{(n-1)}(0)\\
&\diff{^n (f(\alpha)g(\alpha))}{\alpha^n}=\sum_{k=0}^{n}\binom{n}{k}f^{(k)}(\alpha)g^{(n-k)}(\alpha)
\end{aligned}
\end{equation}
perform $n$th-order derivative on both sides of (\ref{eqn:general_he}) to $\alpha$ at $\alpha=0$:
\begin{equation}\label{eqn:general_der_sol}
\begin{aligned}
&\mathbf{A}_0\sum_{k=0}^{n}\binom{n}{k}(\mathbf{x}\otimes\mathbf{1})^{(k)}(0)\circ(\mathbf{1}\otimes\mathbf{x})^{(n-k)}(0)+\\
&n\mathbf{A}_1\sum_{k=0}^{n-1}\binom{n-1}{k}(\mathbf{x}\otimes\mathbf{1})^{(k)}(0)\circ(\mathbf{1}\otimes\mathbf{x})^{(n-1-k)}(0)+\\
&\mathbf{B}_0\mathbf{x}^{(n)}(0)+n\mathbf{B}_1\mathbf{x}^{(n-1)}(0)+\delta_{n,0}\mathbf{C}_0+\delta_{n,1}\mathbf{C}_1=\mathbf{0}
\end{aligned}
\end{equation}

Dividing both sides of (\ref{eqn:general_der_sol}) with $n!$:
\begin{equation}\label{eqn:general_der_solx}
\begin{aligned}
&\mathbf{A}_0\sum_{k=0}^{n}\frac{(\mathbf{x}\otimes\mathbf{1})^{(k)}(0)}{k!}\circ\frac{(\mathbf{1}\otimes\mathbf{x})^{(n-k)}(0)}{(n-k)!}+\\
&\mathbf{A}_1\sum_{k=0}^{n-1}\frac{(\mathbf{x}\otimes\mathbf{1})^{(k)}(0)}{k!}\circ\frac{(\mathbf{1}\otimes\mathbf{x})^{(n-1-k)}(0)}{(n-1-k)!}+\\
&\mathbf{B}_0\frac{\mathbf{x}^{(n)}(0)}{n!}+\mathbf{B}_1\frac{\mathbf{x}^{(n-1)}(0)}{(n-1)!}+\frac{\delta_{n,0}}{n!}\mathbf{C}_0+\frac{\delta_{n,1}}{n!}\mathbf{C}_1=\mathbf{0}
\end{aligned}
\end{equation}

By comparing (\ref{eqn:general_he_sol}) and (\ref{eqn:general_der_solx}) it is concluded that
\begin{equation}\label{eqn:he_der_eqv}
\mathbf{x}[n]=\frac{\mathbf{x}^{(n)}(0)}{n!}
\end{equation}
and the following remarks can be made:
\begin{remark}\label{rem:remark1}
	The HE solution represents the Maclaurin series of $\mathbf{x}(\alpha)$ for problem (\ref{eqn:general_he}).
\end{remark}
\begin{remark}\label{rem:remark2}
	If the $n$th-order truncated series of HE solution $\mathbf{x}_{ps,n}(\alpha)=\sum_{k=0}^{n}\mathbf{x}[k]\alpha^k$ is used to approximate $\mathbf{x}(\alpha)$, then the error is $o(\alpha^n)$.
\end{remark}
\begin{remark}\label{rem:remark3}
	It is shown that the derivation of the HE coefficient equations is equivalent to performing high-order derivatives on both sides of equations at $\alpha=0$. For (\ref{eqn:abstract_he}),
	calculate the derivative of $\alpha$ to both sides, and get
	\begin{equation}\label{eqn:abstract_he_d1}
	\frac{\partial\mathbf{f}}{\partial\mathbf{x}}\diff{\mathbf{x}}{\alpha}=-\pdiff{\mathbf{f}}{\alpha}.
	\end{equation}
	
	Perform $(n-1)$th-order derivative on (\ref{eqn:abstract_he_d1}) w.r.t. $\alpha$:
	\begin{equation}\label{eqn:abstract_he_dn}
	\frac{\partial\mathbf{f}}{\partial\mathbf{x}}\diff{^n\mathbf{x}}{\alpha^n}+\sum_{k=1}^{n-1}\binom{n-1}{k}\diff{^k\partial\mathbf{f}/\partial\mathbf{x}}{\alpha^k}\diff{^{n-k}\mathbf{x}}{\alpha^{n-k}}=-\pdiff{^n\mathbf{f}}{\alpha^n},
	\end{equation}
	note that the second term on the left-hand side has an up-to-$(n-1)$th-order derivative of $\mathbf{x}$ to $\alpha$. Once the derivatives of $(n-1)$th-order or lower are obtained, the $n$th-order derivative $\diff{^n\mathbf{x}}{\alpha^n}$ can be calculated. This also matches the recursive procedure of calculating HE coefficients. Moreover, (\ref{eqn:abstract_he_dn}) reveals that to solve the HE coefficient at any level, one needs to solve linear equations with the same coefficient matrix $\left.\frac{\partial\mathbf{f}}{\partial\mathbf{x}}\right|_{\alpha=0}$ \cite{liu2019solving,liu2020dynamized,yao2019efficient}. This will facilitate the computation because the matrix only needs to be factorized once.
\end{remark}

\section{Partitioned HE (PHE) for contingency Analysis}
\subsection{Interface-based PHE Formulation}
\begin{figure}[htb]
	\centering
	\includegraphics[clip=true,scale=0.1]{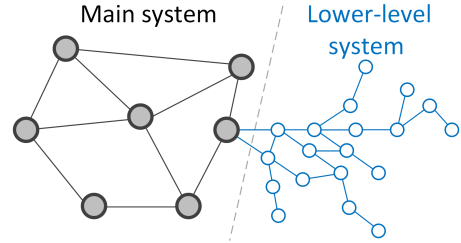}\\
	\caption{Illustration of main system and lower-level systems.}
	\label{fig:system_structure}
\end{figure}

A power system consists of the network and the components connected to the network. The $I_{Li}$ terms in HE formulation (\ref{eqn:power_flow_ctg_nt_pvpq}) for contingency analysis can be generalized as a flexible voltage-current interface compatible with various loads, generation models or lower-level systems. Fig. \ref{fig:system_structure} illustrates the structure of a main system with subsystems connected to it. Each lower-level system is connected to a single node or a limited number of nodes in the main system, and the lower-level systems usually do not have connections to other subsystems to avoid electromagnetic loops.

\begin{figure}[htb]
	\centering
	\includegraphics[clip=true,scale=0.1]{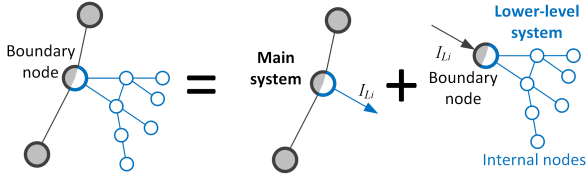}\\
	\caption{Separate modeling of main system and lower-level system(s).}
	\label{fig:system_superposition}
\end{figure}

In this section, we will show that the computation of HE can also be adapted to the inter-area interfaces corresponding to the hierarchical structure of power systems. The nodes on the main system that lower-level systems connect to are called boundary nodes. From the perspective of the main system, the current from the boundary nodes to the lower-level systems is equivalent of the lower-level system, and vice versa: The influence of the main system on the lower-level system can be represented by the voltage and current injection. As Fig. \ref{fig:system_superposition} shows, the coupled system with the main system and the subsystem can be viewed as the superposition of the main system and the lower-level system with injection currents, and then they can be modeled separately with HE. When modeling the lower-level system, the boundary node only acts as a bridge to the main system; all the other components (loads, generators or other shunt components) on the boundary nodes will be modeled with the main system. All other nodes belonging to the lower-level system are called internal nodes.
\begin{equation}\label{eqn:hem_nt_sol_linear_sub}
\begin{aligned}
&\resizebox{1\hsize}{!}{$\left[\phantom{\begin{matrix}a_0\\ \ddots\\a_0\\b_0\\ \ddots\\b_0 \end{matrix}}
	\right.\hspace{-1.5em}
	\underbrace{\begin{array}{c;{2pt/2pt}c;{2pt/2pt}c;{2pt/2pt}c;{2pt/2pt}c}
		-\mathbf{G}_{sii} &  \mathbf{B}_{sii} & \md(\mathbf{P}_{0si}) & -\md(\mathbf{Q}_{0si}) & \begin{matrix}\mathbf{0}\\-\md(\mathbf{F}_{sPV}[0])\end{matrix} \\\hdashline[2pt/2pt]
		-\mathbf{B}_{sii} & -\mathbf{G}_{sii} &-\md(\mathbf{Q}_{0si}) & -\md(\mathbf{P}_{0si}) & \begin{matrix}\mathbf{0}\\-\md(\mathbf{E}_{siPV}[0])\end{matrix} \\\hdashline[2pt/2pt]
		\md(\mathbf{E}_{si}[0]) & -\md(\mathbf{F}_{si}[0]) & \md(\mathbf{C}_{si}[0]) & -\md(\mathbf{D}_{si}[0]) & \mathbf{0} \\\hdashline[2pt/2pt]
		\md(\mathbf{F}_{si}[0]) &  \md(\mathbf{E}_{si}[0]) & \md(\mathbf{D}_{si}[0]) &  \md(\mathbf{C}_{si}[0]) & \mathbf{0}\\\hdashline[2pt/2pt]
		\begin{matrix}\mathbf{0}&\md(\mathbf{C}_{siPV}[0])\end{matrix} & \begin{matrix}\mathbf{0}&\md(\mathbf{D}_{siPV}[0])\end{matrix} & \mathbf{0} & \mathbf{0} & \mathbf{0}  
		\end{array}}_{\textstyle\mathbf{L}_{sii}}\hspace{-1.5em}
	\left.\phantom{\begin{matrix}a_0\\ \ddots\\a_0\\b_0\\ \ddots\\b_0 \end{matrix}}	
	\right]
	\left[\phantom{\begin{matrix}a_0\\a_0\\a_0\\b_0\\b_0\end{matrix}}
	\right.\hspace{-1.5em}
	\underbrace{\begin{array}{c}
		\mathbf{C}_{si}[n]\\\mathbf{D}_{si}[n]\\\mathbf{E}_{si}[n]\\\mathbf{F}_{si}[n]\\\mathbf{Q}_{siPV}[n]
		\end{array}}_{\textstyle\mathbf{x}_{si}[n]}\hspace{-1.5em}
	\left.\phantom{\begin{matrix}a_0\\a_0\\a_0\\b_0\\b_0\end{matrix}}	
	\right]$}\\
&\resizebox{1\hsize}{!}{$=
	-\left[\phantom{\begin{matrix}a_0\\\hdashline[2pt/2pt] b_0\\a_0\\b_0\\ b_0\end{matrix}}
	\right.\hspace{-1.5em}
	\underbrace{\begin{array}{c;{2pt/2pt}c}
		-\mathbf{G}_{sib}&\mathbf{B}_{sib}\\\hdashline[2pt/2pt]
		-\mathbf{B}_{sib} & -\mathbf{G}_{sib}\\\hdashline[2pt/2pt]
		\mathbf{0}&\mathbf{0}\\\hdashline[2pt/2pt]
		\mathbf{0}&\mathbf{0}\\\hdashline[2pt/2pt]
		\mathbf{0}&\mathbf{0}
		\end{array}}_{\textstyle\mathbf{L}_{sib}}\hspace{-1.5em}
	\left.\phantom{\begin{matrix}a_0\\\hdashline[2pt/2pt] b_0\\a_0\\b_0\\ b_0\end{matrix}}
	\right]
	\left[\phantom{\begin{matrix}a_0\\a_0\end{matrix}}
	\right.\hspace{-1.5em}
	\underbrace{\begin{array}{c}
		\mathbf{C}_{sb}[n]\\\mathbf{D}_{sb}[n]
		\end{array}}_{\textstyle\mathbf{x}_{sb}[n]}\hspace{-1.5em}
	\left.\phantom{\begin{matrix}a_0\\a_0\end{matrix}}
	\right]
	+
	\left[\phantom{\begin{matrix}a_0\\a_0\\b_0\\b_0\\a_0\\b_0\\b_0\\a_0\\b_0\\b_0\\a_0\\b_0\\b_0\\a_0\\a_0\end{matrix}}
	\right.\hspace{-1.5em}
	\underbrace{\begin{array}{c}
		\begin{array}{c}
		\Re\displaystyle{\left(\mathbf{I}_{siLPQ}[n]\right)}\\
		\Re\displaystyle{\left(j\sum_{k=1}^{n-1}\mathbf{Q}_{siPV}[k]\circ \mathbf{W}_{siPV}^*[n-k]+\mathbf{I}_{siLPV}[n]\right)}
		\end{array}+\Re\displaystyle{\left(\Delta\mathbf{Y}_{si\cdot}\mathbf{V}_s[n-1]\right)}\\\hdashline[2pt/2pt]
		\begin{array}{c}
		\Im\displaystyle{\left(\mathbf{I}_{siLPQ}[n]\right)}\\
		\Im\displaystyle{\left(j\sum_{k=1}^{n-1}\mathbf{Q}_{siPV}[k]\circ \mathbf{W}_{siPV}^*[n-k]+\mathbf{I}_{siLPV}[n]\right)}
		\end{array}+\Im\displaystyle{\left(\Delta\mathbf{Y}_{si\cdot}\mathbf{V}_s[n-1]\right)}\\\hdashline[2pt/2pt]
		\displaystyle{\Re\left(-\sum_{k=1}^{n-1}\mathbf{W}_{si}[k]\circ \mathbf{V}_{si}[n-k]\right)}\\\hdashline[2pt/2pt]
		\displaystyle{\Im\left(-\sum_{k=1}^{n-1}\mathbf{W}_{si}[k]\circ \mathbf{V}_{si}[n-k]\right)}\\\hdashline[2pt/2pt]
		\displaystyle{-\frac{1}{2}\sum_{k=1}^{n-1}\mathbf{V}_{siPV}[k]\circ \mathbf{V}^*_{siPV}[n-k]}
		\end{array}}_{\textstyle\mathbf{R}_{si}[n]}\hspace{-1.5em}
	\left.\phantom{\begin{matrix}a_0\\a_0\\b_0\\b_0\\a_0\\b_0\\b_0\\a_0\\b_0\\b_0\\a_0\\b_0\\b_0\\a_0\\a_0\end{matrix}}
	\right]
	$}
\end{aligned}
\end{equation}

Assume that the slack bus of the whole system is on the main system. For the main system, the HE formulation is the same as (\ref{eqn:power_flow_ctg_nt_pvpq}) where $I_{Li}$ terms include the current injections to the lower-level systems. The equations of HE coefficients are the same as (\ref{eqn:hem_nt_sol_linear}). The $I_{Li}[n]$ terms depend on the voltage of the boundary nodes as well as the states of the lower-level systems. For the lower-level system, the equations of the HE coefficients have some differences, as (\ref{eqn:hem_nt_sol_linear_sub}) shows. $\mathbf{V}_s$ is the vector containing the voltages of the internal nodes and the boundary nodes of a lower-level system. $\mathbf{V}_{sb}$ is the voltage of boundary nodes, whose real and imaginary parts are $\mathbf{C}_{sb}$ and $\mathbf{D}_{sb}$. $\mathbf{V}_{si}$ is the voltage of internal nodes, whose real and imaginary parts are $\mathbf{C}_{si}$ and $\mathbf{D}_{si}$. $\mathbf{E}_{si}$ and $\mathbf{F}_{si}$ are the real and imaginary parts of $\mathbf{W}_{si}$ i.e., the reciprocal of $\mathbf{V}_{si}$. $\mathbf{Y}_s$ is the admittance matrix with the following blocks, whose rows and columns correspond to the internal and boundary nodes as shown on the subscripts:
\begin{equation}\label{eqn:admittance_mat}
\mathbf{Y}_s=
\left[\begin{array}{c;{2pt/2pt}c}
\mathbf{Y}_{sbb}&\mathbf{Y}_{sbi}\\\hdashline[2pt/2pt]
\mathbf{Y}_{sib}&\mathbf{Y}_{sii}
\end{array}
\right]
\end{equation}
here $\mathbf{Y}_{si\cdot}=[\mathbf{Y}_{sib},\mathbf{Y}_{sii}]$. $\mathbf{G}_{s\cdot}$ and $\mathbf{B}_{s\cdot}$ are the real and imaginary parts of the admittance matrix or its sub-matrices. $\mathbf{I}_{siL}$ is the current to other components on the internal nodes. On the boundary nodes, all the other components have been modeled with the main system, and here on the lower-level system, the boundary nodes act only as bridges absorbing injection currents from the main system and distributing them to the lower-level system. The boundary node does not have any other injection currents. Therefore, for the boundary nodes, the equations of HE coefficients are:
\begin{equation}\label{eqn:hem_nt_sol_linear_bound}
\begin{aligned}
&\left[\phantom{\begin{matrix}a_0\\b_0\end{matrix}}
\right.\hspace{-1.5em}
\underbrace{\begin{array}{c;{2pt/2pt}c;{2pt/2pt}c;{2pt/2pt}c;{2pt/2pt}c}
	-\mathbf{G}_{sii} &  \mathbf{B}_{sii} & \mathbf{0} & \mathbf{0} & \mathbf{0} \\\hdashline[2pt/2pt]
	-\mathbf{B}_{sii} & -\mathbf{G}_{sii} & \mathbf{0} & \mathbf{0} & \mathbf{0}
	\end{array}}_{\textstyle\mathbf{L}_{sbi}}\hspace{-1.5em}
\left.\phantom{\begin{matrix}a_0\\b_0\end{matrix}}	
\right]	
\left[\begin{array}{c}
\mathbf{C}_{si}[n]\\\mathbf{D}_{si}[n]\\\mathbf{E}_{si}[n]\\\mathbf{F}_{si}[n]\\\mathbf{Q}_{siPV}[n]
\end{array}\right]\\
&+\left[\phantom{\begin{matrix}a_0\\b_0\end{matrix}}
\right.\hspace{-1.5em}
\underbrace{\begin{array}{c;{2pt/2pt}c}
	-\mathbf{G}_{sib}&\mathbf{B}_{sib}\\\hdashline[2pt/2pt]
	-\mathbf{B}_{sib} & -\mathbf{G}_{sib}
	\end{array}}_{\textstyle\mathbf{L}_{sbb}}\hspace{-1.5em}
\left.\phantom{\begin{matrix}a_0\\b_0\end{matrix}}
\right]
\left[\begin{array}{c}
\mathbf{C}_{sb}[n]\\\mathbf{D}_{sb}[n]
\end{array}\right]\\
&=-
\left[\begin{array}{c}
\Re\displaystyle{\left(\mathbf{I}_{L}[n]\right)}\\\hdashline[2pt/2pt]
\Im\displaystyle{\left(\mathbf{I}_{L}[n]\right)}
\end{array}
\right]+
\left[\phantom{\begin{matrix}a_0\\b_0\end{matrix}}
\right.\hspace{-1.5em}
\underbrace{\begin{array}{c}
	\Re\displaystyle{\left(\Delta\mathbf{Y}_{sb\cdot}\mathbf{V}_s[n-1]\right)}\\\hdashline[2pt/2pt]
	\Im\displaystyle{\left(\Delta\mathbf{Y}_{sb\cdot}\mathbf{V}_s[n-1]\right)}
	\end{array}}_{\textstyle\mathbf{R}_{sb}[n]}\hspace{-1.5em}
\left.\phantom{\begin{matrix}a_0\\b_0\end{matrix}}
\right]
\end{aligned}
\end{equation}
where on the right-hand side, the $\mathbf{I}_{L}$ term is the current injection from the main system. From (\ref{eqn:hem_nt_sol_linear_sub}) the internal variables can be expressed as:
\begin{equation}\label{eqn:sub_solve_internal}
\mathbf{x}_{si}[n]=-\mathbf{L}_{sii}^{-1}\mathbf{L}_{sib}\mathbf{x}_{sb}[n]+\mathbf{L}_{sii}^{-1}\mathbf{R}_{si}[n],
\end{equation}
and combining with (\ref{eqn:hem_nt_sol_linear_bound}), the internal variables are eliminated:
\begin{equation}\label{eqn:sub_solve_ext}
\begin{aligned}
\phantom{\begin{matrix}a_0\end{matrix}}
\hspace{-1.5em}
\underbrace{(\mathbf{L}_{sbb}-\mathbf{L}_{sbi}\mathbf{L}_{sii}^{-1}\mathbf{L}_{sib})}_{\textstyle\mathbf{L}_{s}}\hspace{-1.5em}
\phantom{\begin{matrix}a_0\end{matrix}}
&\left[\begin{array}{c}
\mathbf{C}_{sb}[n]\\\mathbf{D}_{sb}[n]
\end{array}\right]=\\
-\left[\begin{array}{c}
\Re\displaystyle{\left(\mathbf{I}_{L}[n]\right)}\\\hdashline[2pt/2pt]
\Im\displaystyle{\left(\mathbf{I}_{L}[n]\right)}
\end{array}
\right]&+
\phantom{\begin{matrix}a_0\end{matrix}}
\hspace{-1.2em}
\underbrace{\mathbf{R}_{sb}[n]
	-\mathbf{L}_{sbi}\mathbf{L}_{sii}^{-1}\mathbf{R}_{si}[n].}_{\textstyle\mathbf{R}_{s}[n]}\hspace{-1.5em}
\phantom{\begin{matrix}a_0\end{matrix}}
\end{aligned}
\end{equation}

Eq. (\ref{eqn:sub_solve_ext}) is substituted in the \textit{main-system problem} to eliminate $\mathbf{I}_{L}$. The matrix $\mathbf{L}_{s}$ is merged into the left-hand side matrix $\mathbf{L}$, and $\mathbf{R}_{s}[n]$ is merged into the right-hand side $\mathbf{R}[n]$. Then the main system variables can be solved, which also means that the boundary variables of each lower-level system problem $\mathbf{x}_{sb}[n]$ are obtained. Finally the HE coefficients of the lower-level system $\mathbf{x}_{si}[n]$ are solved from (\ref{eqn:sub_solve_internal}). 
The PHE solution is the same with HE. And PHE does not compromise the numerical robustness of the original HE method. 

\subsection{Comparative complexity analysis of PHE and HE}
A computational complexity analysis can be done to compare the theoretical speed of PHE and HE. To simplify the analysis, assume the entire system consists of a main system with $N_m$ buses and $K$ lower-level systems each with $N_s$ buses. Each lower-level system is connected to the main system through $n_b$ boundary nodes. For the ordinary HE approach (\ref{eqn:hem_nt_sol_linear}) that directly solves the entire system, first we factorize the left-hand-side matrix of (\ref{eqn:hem_nt_sol_linear}), which requires $c_0(N_m+KN_s)^3$ operations, here $c_0$ as well as $c_1$, $c_2$, $c_{Rn}$ below are constants depending on the system structure and bus types. Here we ignore some minor computation costs, e.g. generating matrices. Then for each order $n$ of HE coefficients ($1\leq n\leq N$), assume generating the right-hand-side vector costs $c_{Rn}(N_m+KN_s)$ operations, and using forward and backward substitution to solve the equation costs $c_1(N_m+KN_s)^2$ operations. So the total number of operations of HE approach is:
\begin{equation}\label{eqn:complexity_he}
\begin{aligned}
m_{1}=&c_0(N_m+KN_s)^3+Nc_1(N_m+KN_s)^2\\
&+\sum_{n=1}^{N}c_{Rn}(N_m+KN_s)
\end{aligned}
\end{equation}

Next we analyze the complexity of the PHE approach. At the beginning, for each system $\mathbf{L}_{sii}$ is factorized, and then $\mathbf{L}_{sbi}\mathbf{L}_{sii}^{-1}$ and $\mathbf{L}_{sii}^{-1}\mathbf{L}_{sib}$ are computed, which constitutes $Kc_0N_s^3+2Kn_bc_1N_s^2$ operations for all the lower-level systems. For the main system, factorizing $\mathbf{L}$ costs $c_0N_m^3$ operations. And then for each order $n$ of HE coefficient, generating $\mathbf{R}_{si}[n]$ costs $Kc_{Rn}N_s$ operations, calculating $\mathbf{R}_{s}[n]$ in (\ref{eqn:sub_solve_ext}) costs $Kc_1N_s^2+Kc_2n_bN_s$ operations. On the main system, generating $\mathbf{R}[n]$ costs $c_{Rn}N_m$ operations, and solving the coefficients on the main system costs $c_1N_m^2$. Finally $Kc_2n_bN_s$ operations are needed to calculate HE coefficients on all the lower-level systems based on (\ref{eqn:sub_solve_internal}). 
\begin{equation}\label{eqn:complexity_phe}
\begin{aligned}
m_{2}=&c_0(N_m^3+KN_s^3)+2Kn_bc_1N_s^2+Nc_1(N_m^2+KN_s^2)\\
&+\sum_{n=1}^{N}c_{Rn}(N_m+KN_s)+2NKc_2n_bN_s
\end{aligned}
\end{equation}

Some reference values for the constants are $c_0$=60, $c_1$=40 and $c_2$=9. Considering that $K\geq 1$ and normally $n_b\ll N_m$, $n_b\ll N_s$, we can get $m_2<m_1$, which means that PHE costs fewer operations than HE and thus should be faster than HE. Also we can conjecture that a larger $N_m$, $N_s$, $K$ or a smaller $n_b$ would help PHE gain a larger advantage over HE.

However, some factors that add to computational burden of PHE are not directly reflected in (\ref{eqn:complexity_phe}) but should be considered in practice. First, the $\mathbf{L}_{s}$ may add to non-zero elements of $\mathbf{L}$, which leads to a larger effective $c_0$ and $c_1$ in (\ref{eqn:complexity_phe}). Second, the computational procedures of PHE is more complicated than HE, which means higher overhead for program execution (e.g. memory access and management). Third, the complexity of matrix factorization being $O(N_{(\cdot)}^3)$ ($N_{(\cdot)}$ is the size of a matrix) is based on dense matrix. On sparse matrices, the effective exponent will be lower than 3, which also diminishes the advantage of PHE. Generally, PHE should have more significant advantages over HE on large systems (larger $N_m$, $N_s$) that have multiple lower-level systems (larger $K$) with clear boundaries (smaller $n_b$), so that $m_2$ is significantly lower than $m_1$ and the overheads can be ignored. Fortunately, power systems usually have such traits and thus we can reasonably expect the advantage of PHE over HE on computational efficiency.

%Regarding the computation speed: HE method that directly solves the entire system needs to solve linear equations at the size of the entire system, while the the PHE method solves several smaller equations corresponding to the main system and lower-level systems, which is usually faster. Therefore, especially when the lower-level systems are relatively large, the PHE method will be faster than the ordinary HE method.

\section{Parallel and Distributed contingency analysis}
The parallelism has two aspects. First, the parallel or distributed computation based on network partition is proposed. It should be noted that the generalized inter-area interface enables very compact data transfer, which favors distributed computation. Second, different outages under the same base state can be computed in parallel due to shared data. 
\subsection{Parallel partitioned HE (P${}^\mathrm{2}$HE)}
Section III has demonstrated the computation of PHE based on system partition and voltage-current interfaces among partitions. Such a partitioned scheme enables parallel or distributed computation. The computation is separated into sub-tasks on the main system and lower-level systems, and the data are transferred among the sub-tasks. When there are multiple lower-level systems, the sub-tasks for lower-level systems are independent of each other and can be parallelized. Algorithm 1 shows the procedures of calculating HE solutions with parallel or distributed computation. We assume that each partition of the system corresponds to a process in computer, and the processes either use shared memory or local memory and can communicate with each other. 
\begin{table}[h]
	\centering
	\label{tab:alg2}
	\begin{tabularx}{1.01\linewidth}{XX}
		\toprule[1.5pt]
		\textbf{Algorithm 1.} Parallel computation of HE on partitioned system. \\ \midrule[1pt]
		\textbf{System model:} \\
		\x Admittance matrices of main system $\mathbf{Y}$ and all lower-level systems $\mathbf{Y}_s$.\\
		\x Power injections of the main system $\mathbf{P}_0$, $\mathbf{Q}_0$. \\
		\x Power injections of the lower-level systems $\mathbf{P}_{0s}$, $\mathbf{Q}_{0s}$. \\
		\x Other component models on the main system and lower-level systems.\\ 
		\textbf{Inputs:} \\
		\x Initial states of the main system: $\mathbf{C}[0]$, $\mathbf{D}[0]$, $\mathbf{E}[0]$, $\mathbf{F}[0]$, $\mathbf{Q}_{PV}[0]$.\\
		\x Initial states of lower-level systems: $\mathbf{C}_{si}[0]$, $\mathbf{D}_{si}[0]$, $\mathbf{E}_{si}[0]$, $\mathbf{F}_{si}[0]$,  $\mathbf{Q}_{siPV}[0]$.\\
		\x Initial states of boundary nodes $\mathbf{C}_{sb}[0]$, $\mathbf{D}_{sb}[0]$ (from $\mathbf{C}[0]$, $\mathbf{D}[0]$).\\
		\textbf{Outputs:} HE coefficients for $n=1\cdots N$\\
		\x Main system: $\mathbf{C}[n]$, $\mathbf{D}[n]$, $\mathbf{E}[n]$, $\mathbf{F}[n]$, $\mathbf{Q}_{PV}[n]$.\\
		\x Lower-level systems: $\mathbf{C}_{si}[n]$, $\mathbf{D}_{si}[n]$, $\mathbf{E}_{si}[n]$, $\mathbf{F}_{si}[n]$, $\mathbf{Q}_{siPV}[n]$. \\
		\textbf{Processes:} \\	
		\rowcolor{LightRed} \x Main process (MP): computation tasks of the main system.\\
		\rowcolor{LightBlue}\x Sub-process (SP($s$)): computation tasks of each lower-level system $s$.\\\midrule[1pt]
		\rowcolor{LightBlue}\verb| 1| \textbf{\textit{parallel} foreach} SP($s$) \textbf{do} \\
		\rowcolor{LightBlue}\verb| 2| \x Calculate $\mathbf{L}_{sbb}$, $\mathbf{L}_{sbi}$, $\mathbf{L}_{sib}$, $\mathbf{L}_{sii}$. Factorize $\mathbf{L}_{sii}$.\\
		\rowcolor{LightBlue}\verb| 3| \x Calculate $\mathbf{L}_{s}=\mathbf{L}_{sbb}-\mathbf{L}_{sbi}\mathbf{L}_{sii}^{-1}\mathbf{L}_{sib}$ and \parcommand{send} to the MP.\\
		\rowcolor{LightBlue}\verb| 4| \textbf{end foreach}\\
		\rowcolor{LightRed} \verb| 5| MP \parcommand{receives} all $\mathbf{L}_{s}$ and calculate LHS matrix $\mathbf{L}$ in (\ref{eqn:power_flow_ctg_nt_pvpq}). Factorize $\mathbf{L}$. \\
		\rowcolor{LightBlue}\verb| 6| \textbf{\textit{parallel} foreach} SP($s$) \textbf{do} \\
		\rowcolor{LightBlue}\verb| 7| \x Prepare $\Delta\mathbf{Y}_s$ based on contingency information.\\
		\rowcolor{LightBlue}\verb| 8| \textbf{end foreach}\\
		\rowcolor{LightRed} \verb| 9| MP prepares $\Delta\mathbf{Y}$ based on contingency information.\\
		\verb|10| \textbf{for} $n=1\rightarrow N$ \textbf{do}\\
		\rowcolor{LightBlue}\verb|11| \x \textbf{\textit{parallel} foreach} SP($s$) \textbf{do} \\
		\rowcolor{LightBlue}\verb|12| \x\x Calculate $\mathbf{R}_s[n]$ based on (\ref{eqn:sub_solve_ext}) and \parcommand{send} to MP.\\
		\rowcolor{LightBlue}\verb|13| \x \textbf{end foreach}\\
		\rowcolor{LightRed} \verb|14| \x MP \parcommand{receives} $\mathbf{R}_s[n]$ and calculate RHS term $\mathbf{R}[n]$ in (\ref{eqn:power_flow_ctg_nt_pvpq}).\\
		\rowcolor{LightRed} \verb|15| \x MP solves (\ref{eqn:power_flow_ctg_nt_pvpq}) and obtains $\mathbf{C}[n]$, $\mathbf{D}[n]$, $\mathbf{E}[n]$, $\mathbf{F}[n]$, $\mathbf{Q}_{PV}[n]$.\\
		\rowcolor{LightRed} \verb|16| \x MP \parcommand{sends} $\mathbf{C}_{sb}[n]$, $\mathbf{D}_{sb}[n]$ to each SP($s$).\\
		\rowcolor{LightBlue}\verb|17| \x \textbf{\textit{parallel} foreach} SP($s$) \textbf{do} \\
		\rowcolor{LightBlue}\verb|18| \x\x SP($s$) \parcommand{receives} $\mathbf{C}_{sb}[n]$, $\mathbf{D}_{sb}[n]$ from MP.\\
		\rowcolor{LightBlue}\verb|19| \x\x Solve $\mathbf{C}_{si}[n]$, $\mathbf{D}_{si}[n]$, $\mathbf{E}_{si}[n]$, $\mathbf{F}_{si}[n]$, $\mathbf{Q}_{siPV}[n]$ based on (\ref{eqn:sub_solve_internal}).\\
		\rowcolor{LightBlue}\verb|20| \x \textbf{end foreach}\\
		\verb|21| \textbf{end for}\\
		\bottomrule[1.5pt]
	\end{tabularx}
\end{table}

In Algorithm 1, steps 1-9 prepare the matrices for HE computation and only needs to be done once. Steps 10-21 are the procedures of calculating the HE coefficients from order 1 to $N$. The steps of the main process (MP) and the lower-level system processes (SP) are marked in different colors. It shows that the computation steps on the SPs are independent of each other and thus can be parallelized. For the systems that has many lower-level systems, the P${}^\mathrm{2}$HE method can substantially enhance efficiency compared with the method that directly solves the entire coupled system. Note that unlike some parallel numerical methods based on iterations on the boundaries that usually undermine numerical robustness (i.e., are more likely to be slow or even diverge), the proposed P${}^\mathrm{2}$HE method does not affect the numerical stability at all: It can be verified that the P${}^\mathrm{2}$HE method produces the same result as the HE method that directly solves the entired coupled system. 

The proposed P${}^\mathrm{2}$HE method also favors distributed computation. It can be seen from Algorithm 1 that during the computation, the SPs send $\mathbf{L}_s$ (only once) and $\mathbf{R}_s[n]$ to the MP, and MP sends back $\mathbf{C}_{sb}[n]$ and $\mathbf{D}_{sb}[n]$ to each SP. The data are highly compact and will not cause a high communication burden. Also, such a message-passing scheme does not exchange lots of internal information about each system partition, so the privacy on each partition is well preserved. This is desirable for coordinated analysis among different system owners or operators.

\subsection{Parallelism among contingency analysis tasks}
In practice, contingency screening often involves assessing different contingencies from the same initial state \cite{yao2017risk}. Note that steps 1-5 of Algorithm 1 are dependent only on the initial state and are independent of the contingencies. Therefore, when assessing different contingencies or outages based on the same initial stage, steps 1-5 only needs to be performed once. For the following steps, the procedures of different contingencies are independent of each other and thus can be parallelized.

\section{Test Cases}
%\subsection{Test environment}
%In the following subsections, the tests are performed on a laptop computer with Core i7-6600U CPU (4 logical processors), 8GB DDR4 2133MHz RAM and 256GB SSD. The methods are implemented and tested on Matlab. 

\subsection{Numerical robustness benchmarking on 458-bus system}
This section tests the proposed HE-based contingency analysis method and compare with traditional methods on a reduced North American eastern interconnection (EI) 458-bus system. The system has 40 PV buses, 417 PQ buses and 2,792 branches. We randomly select 30,000 N-25 outage samples. To simulate the variations of load levels in system operations, the load and generation are amplified by a random number uniformly distributed in interval $[1.0,1.2]$. Because the outages may cause the system to separate into islands, and thus new slack bus(es) has to be designated on newly formed islands, which complicates the tests, we exclude such contingency samples. After excluding the cases that cause system separation, there are 29,247 contingency samples. 

We compare the proposed HE approach with damped NR method with different damping factors. The damped NR method for solving an equation $\mathbf{g}(\mathbf{x})=\mathbf{0}$ uses $\Delta\mathbf{x}=-\mu\nabla\mathbf{g}^{-1}\mathbf{g}$ as the correction, where $\mu$ is the damping factor ($\mu=1$ is the ordinary NR method). Table \ref{tab:he_ieee118} shows the statistics of the results given by HE and NR. HE uses $N=10$. Here ``Normal'' means that the method returns a post-contingency solution and the solution is operable. ``Non-practical'' means that the method returns a post-contingency solution but the solution is not a practical one. ``Collapse'' means that the method judges that the system collapses or the method fails to converge. All the solutions obtained by HE approach are operable solutions, so the ``non-practical'' column for HE is not listed. The results show that HE and NR approaches provide consistent results for most cases, but there are substantial chances that NR deliver wrong results. For example, when $\mu=0.1$, there are 11 samples HE gets operable solution but NR does not converge. NR also delivers non-practical solutions for as many as 309 samples. On all the settings in Table \ref{tab:he_ieee118}, NR method delivers wrong results on more than 1\% samples, and decreasing the damping factor $\mu$ does not improve the robustness of NR significantly. 
\begin{table}[htb]
	\centering
	\caption{Contingency analysis statistics under HE and NR methods on 458-bus system}
	\label{tab:he_ieee118}
	\begin{tabularx}{0.8\linewidth}{p{1.2cm}p{1.8cm}@{}*2{>{\centering\arraybackslash}X}@{}}
		\toprule[1pt]
		&               & \multicolumn{2}{c}{HE}            \\\cline{3-4}
		&               & Normal           & Collapse \\\midrule[0.5pt]
		\multirow{4}{*}{NR ($\mu$=0.1)}		&Normal         & 16,204 & 0 \\
		&Non-practical   & 68 & 241 \\
		&Collapse       & 11 & 12,723 \\\cline{2-4}
		&\textbf{Incorrect \%}   & \multicolumn{2}{c}{\textbf{1.09}} \\\midrule[0.5pt]
		\multirow{4}{*}{NR ($\mu$=0.5)}		&Normal         & 16,162 & 0 \\
		&Non-practical   & 89 & 257 \\
		&Collapse       & 32 & 12,707 \\\cline{2-4}
		&\textbf{Incorrect \%}   & \multicolumn{2}{c}{\textbf{1.29}} \\\midrule[0.5pt]
		\multirow{4}{*}{NR ($\mu$=1)}		&Normal         & 16,150 & 0 \\
		&Non-practical   & 96 & 279 \\
		&Collapse       & 37 & 12,685 \\\cline{2-4}
		&\textbf{Incorrect \%}   & \multicolumn{2}{c}{\textbf{1.41}} \\
		\bottomrule[1pt]
	\end{tabularx}
\end{table}

As a remark, the correctness of an NR solution is first verified by substituting it back to the post-contingency power flow equations to check the equation mismatches. If the NR solution satisfies the equations, then a traceback method is applied to check whether the solution is a practical one, which usually has voltage magnitudes around 1 and relatively low voltage angles. The traceback method is an reverse process of the HE-based contingency analysis, i.e. starting from the examined solution at $\alpha=1$ and try to reach back to the pre-contingency state at $\alpha=0$.

\begin{figure}[htb]
	\centering
	\includegraphics[clip=true,scale=0.12]{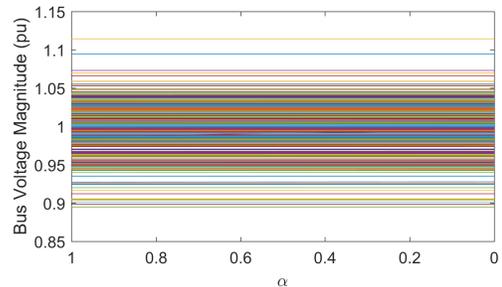}\\
	\caption{Traceback result of a practical solution. $\alpha=1$ corresponds to the examined solution, and the pre-contingency solution is reached at $\alpha=0$.}
	\label{fig:ei_traceback_practical}
\end{figure}

\begin{figure}[htb]
	\centering
	\includegraphics[clip=true,scale=0.12]{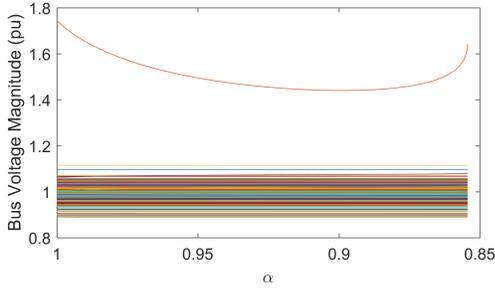}\\
	\caption{Traceback result of a non-practical solution. $\alpha=1$ corresponds to the examined solution. The computation does not reach $\alpha=0$.}
	\label{fig:ei_traceback_nonpractical}
\end{figure}

For a practical operable solution, the traceback method will reach the pre-contingency solution at $\alpha=0$, as Fig. \ref{fig:ei_traceback_practical} shows. If a solution is non-practical, the computation will not reach the pre-contingency solution at $\alpha=0$, like Fig. \ref{fig:ei_traceback_nonpractical} shows. The $>$1\% chance of getting wrong result in NR method may pose a substantial risk to the security analysis of power systems, especially considering that contingency analysis is an important and routine task. In contrast, the HE approach provides correct results and thus has much better credibility than NR method. 

\begin{figure}[htb]
	\centering
	\includegraphics[clip=true,scale=0.08]{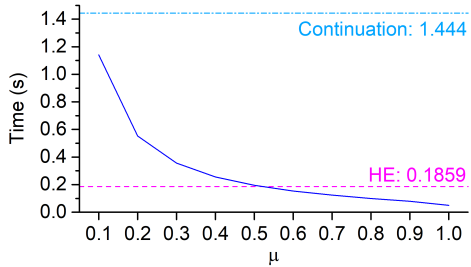}\\
	\caption{Average computation time of HE, NR (with various damping factors $\mu$) and continuation methods for each contingency sample. }
	\label{fig:118_speed}
\end{figure}

Fig. \ref{fig:118_speed} shows the average computation time of NR, HE and continuation methods the 29,274 contingency samples. Results show that the ordinary NR method ($\mu$=1) is the fastest, but also has the highest change to deliver incorrect results. The average time consumption of HE is about 3.73 times that of the ordinary NR method and is about the same as damped NR method with $\mu$=0.5. The continuation method is much slower than the HE and NR approaches. Although HE is slower than the some NR methods, the HE method is desirable because it provides credible analysis results. Compared with  the continuation method, HE has significant advantage in computational efficiency.

\subsection{Contingency analysis on synthetic 118-bus + 43-bus system}
\begin{figure}[htb]
	\centering
	\includegraphics[clip=true,scale=0.1]{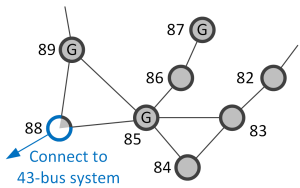}\\
	\caption{Partial network of 118-bus system at the connection point. ``G'' means generation bus.}
	\label{fig:118_local_structure}
\end{figure}

This test case demonstrates the outage analysis using the system partition and PHE method. The system is synthesized by connecting the bus 1 of the 43-bus system \cite{iwamoto1981load} to the bus 88 of the 118-bus system. The 118-bus system has meshed topology, which represents the high-voltage main system, while the 43-bus system with radial topology represents the lower-level system. The original 43-bus test system is very ill-conditioned, and we reduce its load by 60\% in this case to ensure the existence of base-case power flow solution. Fig. \ref{fig:118_local_structure} shows the network structure near the connection point.
Traditionally, the lower-level systems are often simplified as base-state power injections. At the base state, the power injection to the 43-bus system is $S_{88}=3.3791+j0.2599$. We demonstrate the effect on the lower-level system when outage occurs by comparing the post-contingency states with the full lower-level system model and those with equivalent power injection. HE is used to solve the states after the loss of branch 88-89 and branch 85-89, respectively. The results in Table \ref{tab:post_cont_118_43} show that the equivalent power injection may be too optimistic or even deliver totally different results, and thus proves that contingency analysis with full lower-level system model is necessary. 
\begin{table}[htb]
	\centering
	\caption{Post-contingency voltage on bus 88.}
	\label{tab:post_cont_118_43}
	\begin{tabularx}{0.8\linewidth}{p{2.0cm}@{}*2{>{\centering\arraybackslash}X}@{}}
		\toprule[1pt]
		Contingency    & Full lower-level system model & Equivalent injection\\\midrule[0.5pt]
		85-89          & $0.9374\angle-23.23^\circ$ & $0.9404\angle-23.24^\circ$\\
		88-89          & System collapsed           & $0.7507\angle-62.44^\circ$\\
		\bottomrule[1pt]
	\end{tabularx}
\end{table}

We further expand the scale of the synthetic system by connecting 7$\times$43-bus systems to buses 15, 41, 49, 69, 88, 89, and 96 of the 118-bus system. The entire system has 419 buses. Each 43-bus system is connected to the 118-bus system by a line with serial impedance $z=0.002+j0.02$. The 118-bus system is regarded as the main system and the 7$\times$43-bus systems are treated as lower-level systems. The PHE and P${}^\mathrm{2}$HE methods are tested and compared with the HE method directly solving the entire system (namely HE). The P${}^\mathrm{2}$HE method is implemented by using the Matlab \verb|parfor| syntax. The methods are used to perform N-1 analysis of the system. Table \ref{tab:comp_time_118_43} shows the computation time of each method to finish N-1 screening (177 contingencies) of the synthetic system. Results show that P${}^\mathrm{2}$HE is slightly faster than the HE method, while PHE is slower than HE. The performance of P${}^\mathrm{2}$HE is affected by the overhead of \verb|parfor|: by profiling the time consumption on each part of the program, we estimate the theoretical computation time is around 8 s. Since the lower-level systems in this case are relatively small, the PHE method does not show an advantage in computation speed.
\begin{table}[htb]
	\centering
	\caption{Computation time on 419-bus system.}
	\label{tab:comp_time_118_43}
	\begin{tabularx}{0.6\linewidth}{>{\centering\arraybackslash}p{1.5cm}@{}*1{>{\centering\arraybackslash}X}@{}}
		\toprule[1pt]
		Method  & Computation time (s)      \\\midrule[0.5pt]
		HE      & 14.245                    \\
		PHE     & 18.040                    \\
		P${}^\mathrm{2}$HE    & 13.787                    \\
		\bottomrule[1pt]
	\end{tabularx}
\end{table}

\subsection{Contingency analysis on synthetic large-scale system}
A large-scale synthetic system is created by connecting 9$\times$2383-bus Polish systems. The whole system has 21,447 buses, 26,078 branches, 2934 PV buses, 18,514 PQ buses, and 756 ZIP loads. The connecting branches are listed in Table \ref{tab:polish_connect}.

\begin{table}[htb]
	\centering
	\caption{Connections between main system and lower-level systems.}
	\label{tab:polish_connect}
	\begin{tabularx}{0.8\linewidth}{>{\centering\arraybackslash}p{1.5cm}@{}*1{>{\centering\arraybackslash}X}@{}}
		\toprule[1pt]
		Sub-system & Branches (Main bus\#-Sub bus\#) \\\midrule[0.5pt]
		2          & 18-18                     \\
		3          & 448-445,474-475           \\
		4          & 2254-2248, 2247-2250      \\
		5          & 1089-1092, 1100-1095      \\
		6          & 673-665                   \\
		7          & 1354-1356, 1544-1547, 738-739 \\
		8          & 1100-1092, 1089-1095                     \\
		9          & 146-148           \\
		\bottomrule[1pt]
	\end{tabularx}
\end{table}
100 randomly selected N-1 contingencies are tested with HE, PHE and P${}^\mathrm{2}$HE methods on the synthetic system, and the computation time of each method is listed in Table \ref{tab:comp_time_polish}. Results show that the PHE method is significantly faster than HE, which verifies its advantage when analyzing large-scale systems. And P${}^\mathrm{2}$HE can further accelerate the computation by making use of parallelism.

\begin{table}[htb]
	\centering
	\caption{Computation time of 100 N-1 analysis on 21447-bus system.}
	\label{tab:comp_time_polish}
	\begin{tabularx}{0.6\linewidth}{>{\centering\arraybackslash}p{1.5cm}@{}*1{>{\centering\arraybackslash}X}@{}}
		\toprule[1pt]
		Method  & Computation time (s)      \\\midrule[0.5pt]
		HE      & 414.65                    \\
		PHE     & 118.84                    \\
		P${}^\mathrm{2}$HE    & 48.64                    \\
		\bottomrule[1pt]
	\end{tabularx}
\end{table}
\section{Conclusion}
This paper presents steady-state contingency analysis approaches based on holomorphic embedding (HE). The paper first presents the HE formulation and algorithm for contingency analysis, and then summarizes some desirable properties of HE through theoretical analysis, including that the linear equations of HE coefficients have a constant coefficient matrix, which lays the foundation for the proposed methods. Considering the ubiquitous structure of power systems with main system and lower-level systems, the partitioned HE (PHE) method is extended from the basic HE formulation based on the generic voltage-current interface on the boundaries. By partitioning the system and computation in PHE, the computation burden can be substantially reduced. Moreover, the PHE formulation has lower-level system computation tasks independent of each other, which is parallelized as a parallel PHE (P${}^\mathrm{2}$HE) method. The proposed HE method for contingency analysis is compared with the traditional Newton-Raphson on a 458-bus system and demonstrates its advantage in numerical robustness. The PHE and P${}^\mathrm{2}$HE approaches are tested on synthetic IEEE 118-bus + 43-bus system and a 419-bus system and a large-scale 21,447-bus system. The results show that the partitioned and parallel HE methods can significantly accelerate the contingency analysis on large-scale multi-area systems. Note that unlike the traditional participation methods requiring numerical iterations on the boundaries, the proposed PHE and P${}^\mathrm{2}$HE methods are completely equivalent to the basic HE method and the computational robustness is not compromised. The proposed methods have promising potentials for various computational tasks, such as the security analysis of multi-area systems and transmission-distribution co-analysis.

% or


%\appendices

%\section{}

%\section*{Acknowledgment}

\ifCLASSOPTIONcaptionsoff
  \newpage
\fi



%\bibliographystyle{IEEEtran}

\bibliographystyle{IEEEtran}
\bibliography{bib/refs}


%\end{IEEEbiography}


%\newpage



%\vfill


\end{document}